\documentclass[fleqn,12pt,twoside]{article}
\usepackage{espcrc1}

\usepackage{graphicx}
\usepackage[figuresright]{rotating}

\newcommand{\AmS}{{\protect\the\textfont2
  A\kern-.1667em\lower.5ex\hbox{M}\kern-.125emS}}

\title{Pion re-scattering in $\pi^0$ production near threshold}

\author{V. Malafaia\address[CFIF]{CFIF and Department of Physics, Instituto Superior T\'ecnico, \\1049-001 Lisboa, Portugal}
        \thanks{Work supported by FCT under the grant SFRD/BD/4876/2001.}
        and 
        M. T. Pe\~na\addressmark\thanks{Work supported by FCT under the grant CERN/FIS/43709/2001.}}

\begin{document}

\maketitle

\begin{abstract}
We compare the most used approximations to calculate the contribution
from pion re-scattering
mechanisms to pion production from proton-proton scattering, near threshold.
Our calculation employs a  pseudo-vector coupling for the $\pi NN$
vertex and realistic amplitudes for the $\pi N$ re-scattering and the $NN$ transitions.
\end{abstract}

\section{Introduction}
Due to isospin suppression of the otherwise dominant low-energy isovector Weinberg-Tomosawa  $\pi$N term in
the re-scattering diagram,
and the negative interference between the remaining impulse and isoscalar pion re-scattering terms\cite{Park}, 
the empirical data  for $p p \rightarrow p p \pi^0$ near
threshold is only explained by phenomenological non-pionic, short-range,
two-body mechanisms\cite{Riska}, or alternatively by an off-shell extrapolation of the $\pi N$ amplitude\cite{Oset}. 
The high quality data for $p p \rightarrow p p \pi^0$ becomes therefore specially interesting by
allowing to establish the importance of the short-range mechanisms, 
which are difficult to constrain: they
are subsumed in the low energy constants appearing also in the two- and three-nucleon potentials,
but so far these constants have not yet been precisely determined\cite{Epel}.

Nevertheless, any conclusion to be read off from the $\pi^0$ production data is
only meaningful and quantitative provided that the calculation of the re-scattering diagram\cite{Hanh1}
is under absolute control. This is  not the case in calculations performed till now, where approximations at the level of the 
energy integration of the corresponding Feynman diagram were considered. Namely, non-relativistic reductions of the 
 Feynman $\pi$-exchange diagram were done, through choices for the energy of the exchanged pion, both for its propagator
and for the $\pi N-\pi N$  amplitude at the re-scattering vertex.

This work investigates the validity of the mostly used approximations.
It generalizes the work of Refs.\cite{Hanh1} which considers a toy model for
scalar particles and interactions. Our calculation employs a physical model for nucleons and pions, with 
a pseudo-vector coupling for the $\pi N N$ vertex; the $\chi$Pt $\pi N - \pi N$ re-scattering amplitude\cite{Park} 
and the Bonn B potential for the nucleon-nucleon interaction in the final state.

\section{From the Feynman diagram to DWBA}

Starting with the Feynman diagram, the energy integration in the energy $Q'_0$ of the exchanged pion was done after a partial 
fraction decomposition, which isolates the poles of the pion propagator. This way we separated the DWBA amplitude from the stretched boxes 
(where two exchanged particles may be in flight simultaneously) contribution. Then,
\begin{equation}
\mathcal{M}_{DWBA}=\frac{1}{2} \int \frac{d^3 q'}{\left(2 \pi \right)^3 } \left[ V \left( \omega_{\pi} \right)  G_{\pi}
\frac{\left( \omega_1-\omega_2\right)+E_{\pi}}{\omega_{\pi}} \right]
\frac{1}{\left(E_1+E_2-E_{\pi}-\omega_1-\omega_2 \right)} T_{NN}^{FSI} \label{dwbafsi}
\end{equation}
is the exact expression for the DWBA amplitude, where $T$ stands for the transition-matrix of the final-state interaction and 
$V \left( \omega_\pi \right)$ is the product of the $\pi N - \pi N$ amplitude with the $\pi N N$ vertex.
The pion propagator turned to be exactly given by
\begin{equation}
G_{\pi}=\frac{1}{\left[\frac{\omega_1-\omega_2}{2}+\frac{E_{\pi}}{2} \right]^2- \left[\left(
E_{tot}-E-\frac{E_{\pi}}{2}\right)-\frac{\omega_1+\omega_2}{2}-\omega_{\pi} \right]^2},
\end{equation}
Thus, Eq.(\ref{dwbafsi}) is taken as the reference result, in order to investigate the effect of the most common approximations 
hitherto used for the pion propagator in a DWBA final-state calculation. 
These approximations correspond to the following three different choices:
\begin{equation}
\begin{array}{llllll}
G_\pi^{fk} & \equiv & \frac{1}{\left( \frac{m_{\pi}}{2}\right)^ 2-\omega_{\pi}^2 } & &\hspace{1.5cm} &{\rm fixed} \hspace{0.1in}
{\rm  kinematics}\\
G_\pi^{on} & \equiv & \frac{1}{\left( E-\omega_2 \right)^ 2-\omega_{\pi}^2 } & & &{\rm on-shell} \\
G_\pi^{st} & \equiv & - \frac{1}{\omega_{\pi}^2 } & & &{\rm static} \hspace{0.1in} {\rm approximation} 
\end{array}
\label{gappfsi}
\end{equation}
The ``fixed kinematics'' approximation ``freezes'' the energy for the exchanged pion as the difference between the two
on-mass-shell energies of the nucleon, before and after re-scattering, at $m_\pi/2$, its value at threshold; 
the ``on-shell approximation''
considers for the energy of the pion the difference between the same on-mass-shell energies of the nucleon,
but it varies with the energy available for the production process,
and is not frozen at the threshold point.
Naturally, the two approximations coincide at threshold and the deviation between them increases with energy.
The ``static approximation'' makes the pion exchange instantaneous and therefore neglects energy transfer.
It is part of the usual non-relativistic approximation for exchange Feynman diagrams.

\section{Results and Conclusions} 

\begin{figure}
\centering
\includegraphics[width=16cm,keepaspectratio]{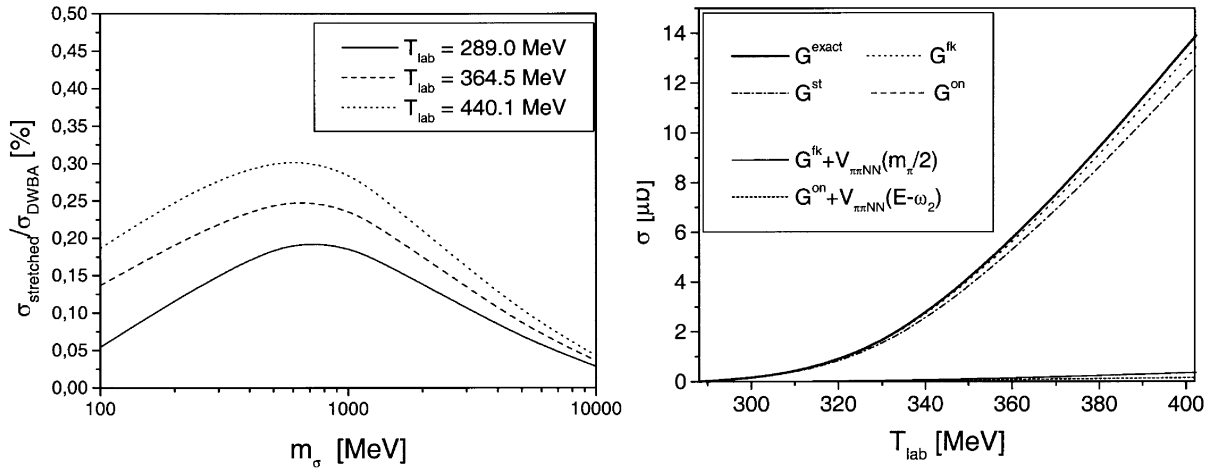}
\caption{Left: The stretched boxes compared to DWBA (FSI) for the cross section as a function of the mass 
of the exchanged particle. Right:  Effect of the choices for the energy of the exchanged pion. The nucleon-nucleon interaction is the Bonn B potential.} 
\label{segfsi}
\end{figure}

For the toy model case of scalar particles and scalar interactions,
we verified the dominance of the DWBA amplitude over the stretched boxes amplitude, as obtained in Ref.\cite{Hanh1}. 
Furthermore, our calculations show that the DWBA keeps being dominant, as seen on the left panel of  Fig.\ref{segfsi}, 
even in a more pronounced way for the physical case considered here.
We found also that within the scalar
toy model all the approximations for the
pion energy taken at the $\pi N - \pi N$ re-scattering amplitude
overestimate the cross section, as found in Refs.\cite{Hanh1}.
However, on the right panel of Fig.\ref{segfsi}, where we describe the nucleon-nucleon interaction by the Bonn B potential T-matrix, 
only the on-shell $G^{on}$ approximation to the exact pion propagator
gives a larger cross section (on the right panel of Fig.\ref{segfsi} this curve practically coincides with the reference result).
The choice for the energy in the pion propagator is not very decisive,
justifying the usual low-energy static or instantaneous approximation for exchange Feynman diagrams. However, 
the choice for the energy of the exchanged pion for the re-scattering vertex,
as prescribed from the 4-dimensional Feynman diagram, is crucial for the cross section strength. As shown on the right panel of 
Fig. \ref{segfsi}, the fixed kinematics and the on-shell prescriptions for the re-scattering vertex are not suitable
for combination with an energy independent $NN$ potential. 

Both for the final- and initial-state interaction, our results show that
 the DWBA formalism is quite adequate at threshold since this part of the full amplitude is clearly
dominant over the stretched boxes. Also,the effect of the usual choices for the $\pi N$ re-scattering energy, which is not 
fixed by a non-relativistic formalism, can be significant.  

In order to extract quantitatively from the data the unknown strength of the short-range
mechanisms for the process $pp \rightarrow pp \pi^0$, one has to use
for the pion  exchange diagram the reference amplitudes. The latter were obtained
as non-relativistic reductions of the corresponding Feynman diagrams for the final- and initial-state interactions, respectively.

\end{document}